\documentclass[twocolumn]{aastex6}

\pdfoutput=1

\newcommand{\mearth}{$M_{\oplus}$}
\newcommand{\rearth}{$R_{\oplus}$}
\newcommand{\mjup}{$M_J$}
\newcommand{\rjup}{$R_J$}
\newcommand{\msun}{$M_{\odot}$}
\newcommand{\rsun}{$R_{\odot}$}
\newcommand{\Teff}{$T_{\mathrm{eff}}$}

\newcommand{\packagename}{{\tt Forecaster}}
\newcommand{\packagelink}{https://github.com/chenjj2/forecaster}
\newcommand{\mytable} {Table~}
\newcommand{\myeqn} {Equation~}
\newcommand{\myfig} {Figure~}
\newcommand{\me}{M_{\oplus}}
\newcommand{\re}{R_{\oplus}}
\newcommand{\normal} {\mathcal{N}}
\newcommand{\uniform} {\mathcal{U}}
\newcommand{\powerconst}{C}
\newcommand{\hyper}{ \boldsymbol{\Theta}_{\mathrm{hyper}} }
\newcommand{\local}{ \boldsymbol{\Theta}_{\mathrm{local}} }
\newcommand{\offset} {\mathcal{C}}
\newcommand{\slope} {S}
\newcommand{\scatter} {\sigma_\mathcal{R}}
\newcommand{\transition} {T}
\newcommand{\logm} {\mathcal{M}}
\newcommand{\logmeq} {\log_{10}(\omob/\me)}
\newcommand{\mt} {\mathcal{M}_t}
\newcommand{\mob} {\mathcal{M}_{ob}}
\newcommand{\merr} {\Delta \mathcal{M}_{ob}}
\newcommand{\logr} {\mathcal{R}}
\newcommand{\logreq} {\log_{10}(\orob/\re)}
\newcommand{\rt} {\mathcal{R}_t}
\newcommand{\rob} {\mathcal{R}_{ob}}
\newcommand{\rerr} {\Delta \mathcal{R}_{ob}}
\newcommand{\omob} {M}
\newcommand{\omt} {M_t}
\newcommand{\omerr} {\Delta M}
\newcommand{\orob} {R}

\newcommand{\orerr} {\Delta R}
\newcommand{\errup} {\Delta_+}
\newcommand{\errdown} {\Delta_-}
\newcommand{\errmean} {\frac{1}{2} (\errup+\errdown)}
\newcommand{\ith} {^{(i)}}
\newcommand{\jth} {^{(j)}}
\newcommand{\kth} {^{(j+1)}}
\newcommand{\fst} {^{(1)}}
\newcommand{\snd} {^{(2)}}
\newcommand{\trd} {^{(3)}}
\newcommand{\fth} {^{(4)}}
\newcommand{\onetofour} {^{(1-4)}}
\newcommand{\onetothree} {^{(1-3)}}
\newcommand{\onetotwo} {^{(1-2)}}
\newcommand{\ninitial} {5}
\newcommand{\ncontinue} {10}
\newcommand{\logdiff} {34.16}
\newcommand{\prob} {p}

\newcommand{\lcorr} {\ell_{corr}}
\newcommand{\marr} { \{ M \ith, i=1,2,...,n \} }
\newcommand{\rarr} { \{  R \ith, i=1,2,...,n \}}
\newcommand{\mgrid} { \{ M_{grid} \jth, j=1,2,...,m \}}
\newcommand{\pgrid} { \{ \prob_{grid} \jth, j=1,2,...,m\} }

\newcommand{\Ndata}{316}
\newcommand{\Nxdata}{632} 

\newcommand{\terr}{Terran}
\newcommand{\Terr}{Terran}
\newcommand{\terrs}{Terrans}

\newcommand{\nept}{Neptunian}
\newcommand{\Nept}{Neptunian}
\newcommand{\nepts}{Neptunians}

\newcommand{\jovi}{Jovian}
\newcommand{\Jovi}{Jovian}
\newcommand{\jovis}{Jovians}

\newcommand{\stel}{Stellar}
\newcommand{\Stel}{Stellar}

\newcommand{\TESS}{\textit{TESS}}
\newcommand{\JWST}{\textit{JWST}}
\newcommand{\CHEOPS}{\textit{CHEOPS}}
\newcommand{\HDST}{\textit{HDST}}

\usepackage{amsmath}
\usepackage{amsfonts}
\usepackage{amssymb}
\usepackage{wasysym}

\begin{document}


\title{Probabilistic Forecasting of the Masses and Radii of Other Worlds}


\author{Jingjing Chen\altaffilmark{1} and David Kipping\altaffilmark{1}}
\affil{Department of Astronomy\\
Columbia University \\
550 W 120th St. \\
New York, NY 10027, USA}


\altaffiltext{1}{jchen@astro.columbia.edu}

\begin{abstract}

Mass and radius are two of the most fundamental properties of an astronomical
object. Increasingly, new planet discoveries are being announced with a 
measurement of one of these terms, but not both. This has led to a growing need
to forecast the missing quantity using the other, especially when 
predicting the detectability of certain follow-up observations. We present 
am unbiased forecasting model built upon a probabilistic mass-radius relation 
conditioned on a sample of \Ndata\ well-constrained objects. Our
publicly available code, \href{\packagelink}{\packagename}, accounts for
observational errors, hyper-parameter uncertainties and the intrinsic dispersions 
observed in the calibration sample. By conditioning our model upon a sample
spanning dwarf planets to late-type stars, \href{\packagelink}{\packagename}
can predict the mass (or radius) from the radius (or mass) for objects covering
nine orders-of-magnitude in mass. Classification is naturally performed by
our model, which uses four classes we label as \terr\ worlds, \nept\ worlds,
\jovi\ worlds and stars. Our classification identifies dwarf planets as
merely low-mass \terrs\ (like the Earth), and brown dwarfs as merely high-mass 
\jovis\ (like Jupiter). We detect a transition in the mass-radius relation at 
$2.0_{-0.6}^{+0.7}$\,\mearth, which we associate with the divide between solid,
\terr\ worlds and \nept\ worlds. This independent analysis adds further weight 
to the emerging consensus that rocky Super-Earths represent a narrower region 
of parameter space than originally thought. Effectively, then, the Earth is the
Super-Earth we have been looking for.

\end{abstract}




\section{INTRODUCTION}
\label{sec:intro}

Over the last two decades, astronomers have discovered thousands of
extrasolar worlds (see \href{www.exoplanets.org}{exoplanets.org};
\citealt{2014PASP..126..827H}), filling in the parameter space from
Moon-sized planets (e.g. \citealt{2013Natur.494..452B}) to brown dwarfs
many times more massive than Jupiter (e.g. \citealt{2008A&A...491..889D}).
Over 98\% of these detections have come from radial velocity, microlensing
or transit surveys, yet each of these methods only directly measures the
mass ($M$) \textit{or} radius ($R$) of planet, not both\footnote{Except for the
rare cases of systems displaying invertible transit timing variations.}.

This leads to the common situation where it is necessary to forecast what
the missing quantity is based on the other. A typical case would be when
one needs to predict the detectability of a potentially observable effect
for a resource-intensive, time-competitive observing facility, which in
some way depends upon the missing quantity. For example, the \TESS\ mission
\citep{2014SPIE.9143E..20R} will soon start detecting hundreds, possibly
thousands, of nearby transiting planets for which the radius, but not the 
mass, will be measured. Planets with radii consistent with Super-Earths will
be of great interest for follow-up and so radial velocity facilities will need
to forecast the detectability, which is proportional to the planet mass, for 
each case. Vice versa, the \CHEOPS\ mission \citep{2013EPJWC..4703005B} will 
try to detect the transits of planets discovered with radial velocities,
necessitating a forecast of the radius based upon the mass.

In those two examples, the objective was to forecast the missing quantity in
order to predict the feasibility of actually measuring it. However, 
the value of forecasting the mass/radius for the purposes of predicting
detectability extends beyond this. As another example, exoplanet transit 
spectroscopy is expected to be a major function of the upcoming JWST mission
\citep{2009ASSP...10..123S}. At the first-order level, the detectability of
an exoplanet atmosphere is proportional to the scale height, $H$, which in turn
is proportional to $1/g \propto R^2/M$. Given the limited supply of cryogen 
onboard \JWST, discoveries of future Earth-analog candidates may be found with 
insufficient time to reasonably schedule a radial velocity campaign first (if 
even detectable at all). Therefore, there will likely be a critical need to 
accurately forecast the scale height of new planet discoveries from just either
the mass or (more likely) the radius.

Forecasting the mass/radius of an object, based upon the other quantity
is most obviously performed using a mass-radius (MR) relation. Such relations
are known to display sharp changes at specific locations, such as the 
transition from brown dwarfs to hydrogen burning stars (e.g. see 
\citealt{2015arXiv150605097H}). These transition points can be thought of as 
bounding a set of classes of astronomical objects, where the classes are 
categorized using the features of the inferred MR relation. In this
case then, it is apparent that inference of the MR relation enables both 
classification and forecasting.

Classification is more than a taxonomical enterprise, it can have dramatic
implications in astronomy. Perhaps the most famous example of classification in
astronomy is the Hertzsprung-Russell (HR) diagram \citep{1909AN....179..373H,
1914PA.....22..275R} for luminosity versus effective temperature, which 
revealed the distinct regimes of stellar evolution. A common concern in 
classification is that the very large number of possible features against which
to frame the problem can be overwhelming. Mass and radius, though, are not 
random and arbitrary choices for framing such a problem. Rather, they are two 
of the most fundamental quantities describing any object in the cosmos and 
indeed represent two of the seven base quantities in the International System 
of Units (SI).

The value of classification extends beyond guiding physical understanding, it 
even affects the design of future instrumentation. As an example, the boundary
between terrestrial planets and Neptune-like planets represents a 
truncation of the largest allowed habitable Earth-like body. The location of
this boundary strongly affects estimates of the occurrence rate of Earth-like 
planets ($\eta_{\oplus}$) and thus in-turn the design requirements of future 
missions needed to characterize such planets \citep{2015arXiv150704779D}. To 
illustrate this, using the occurrence rate posteriors of 
\citet{2014ApJ...795...64F}, 
$\eta_{\oplus}$ decreases by 42\% when altering the
definition of Earth-analogs from $R<2.0$\,\rearth\ to $R<1.5$\,\rearth. 
In order to maintain the same exoEarth yield for the proposed \HDST\ mission,
this change corresponds to a 27\% increase in the required mirror diameter
(using yield equation in \S3.5.4 of \citealt{2015arXiv150704779D}).

We therefore argue that both forecasting and classification using the
masses and radii of astronomical bodies will, at the very least, be of great
utility for present/future missions and may also provide meaningful insights
to guide our interpretation of these objects. Accordingly, the primary
objective of this work is to build a statistically rigorous and empirically 
calibrated model

\begin{itemize}
\item[{\small$\blacktriangleright$}]
to forecast the mass/radius of an astronomical object based upon a
measurement of the other, and
\item[{\small$\blacktriangleright$}]
for the classification of astronomical bodies based upon their observed 
masses and/or radii.
\end{itemize}

The layout of this paper is as follows. In Section~\ref{sec:model}, we
outline our model for the MR relation, which is used enable forecasting
and classification. In Section~\ref{sec:analysis}, we describe the
regression algorithm used to conduct Bayesian parameter estimation of
our model parameters. The results, in terms of both classification and
forecasting are discussed separately in Sections~\ref{sec:classification}
\& \ref{sec:forecasting}. We summarize the main findings of our work
in Section~\ref{sec:discussion}.

\section{MODEL}
\label{sec:model}

\subsection{Choosing a Model}
\label{sub:philosophy}

We begin by describing the rationale behind the model used in this work.
As discussed in Section~\ref{sec:intro} (and demonstrated later in 
Section~\ref{sec:analysis}), the two primary goals of this paper are both achievable
through the use of a MR relation and this defines the approach in this work.
Broadly speaking, such a relation can be cast as either a parametric (e.g. a
polynomial) or non-parametric model (e.g. a nearest neighbor algorithm).

Parametric models, in particular power-laws, have a long been popular for
modeling the MR relation with many examples even in the recent literature (e.g.
\citealt{2006Icar..181..545V,2013ApJ...768...14W,2015arXiv150605097H,
2015arXiv150407557W,2016ApJ...819..127Z}). In our case, we note that such 
models are more straightforward for hierarchical Bayesian modeling (which we 
argue to be necessary later), since they allow for a simple prescription of the
Bayesian network. Moreover, based on those earlier cited works, power-laws 
ostensbily do an excellent job of describing the data and the greater
flexibility afforded by non-parametric methods is not necessary. Accordingly, 
we adopt the power-law prescription in this work.

As noted earlier, the use of power-laws to describe the MR relation is common
in the literature. However, many of the assumptions and model details in these
previous implementations would make forecasts based upon these relations
problematic. We identify three key aspects of the model proposed in this
work with differentiate our work from previous studies.

\textbf{[1] Largest data range:}
Inferences of the MR relation often censor the available data to a specific 
subset of parameter space (for example \citealt{2015arXiv150407557W} consider 
the $R<8$\,\rearth\ exoplanets). Whilst it is inevitable that certain 
subjective choices will be made by those analyzing the MR relation, a more 
physically-motivated choice for the parameter limits can be established. 
Ideally, this range should be as large as possible such that forecasting is 
unlikely to encounter the extrema, leading to truncation errors. A natural 
lower bound is an object with sufficient mass to achieve hydrostatic 
equilibrium leading to a nearly spherical shape and thus a well-defined radius
(a planemo), which would encompass dwarf planets. As an upper bound, late-type
stars take longer than a Hubble time to leave the main-sequence a so should
exhibit a relatively tight trend between mass and radius.

\textbf{[2] Fitted transitions:}
As a by-product of using such a wide mass range, several transitional regions 
are traversed where the MR relation exhibits sharp changes. For example, the 
onset of hydrogen burning leads to a dramatic change in the MR relation versus
brown dwarfs \citep{2015arXiv150605097H}. In previous works, such transitional
points are often held as fixed, assumed locations (e.g. 
\citealt{2014ApJ...783L...6W} assume a physically motivated, but not freely 
inferred, break at 1.5\,$R_{\oplus}$). In contrast, we here seek to make a more 
agnostic, data-driven inference without imposing any assumed transition points 
from theory or previous data-driven inferences. In this way, the uncertainty in 
these transitions is propagated into the inference of all other parameters 
defining our model, leading to more robust uncertainty estimates for both 
forecasting and classification. Accordingly, in this work, the MR relation is 
described by a broken power-law with freely fitted transition points (in addition 
to the other parameters).

\textbf{[3] Probabilistic modeling:}
Whilst mass can be considered to be the primary influence on the size of an 
object, many second-order terms will also play a role. As an example, rocky
planets of the same mass but different core mass fractions will exhibit 
distinct radii \citep{2016ApJ...819..127Z}. When viewed in the MR plane then, a
particular choice of mass will not correspond to a singular radius value. 
Rather, a distribution of radii is expected, as a consequence of the numerous 
hidden second-order effects influencing the size. Statistically speaking then,
the MR relation is expected to be {\it probabilistic}, rather than 
deterministic. A probabilistic model fundamentally relaxes the assumption that
the underlying model (in our case a broken power-law) is the ``correct'' or 
``true'' description of the data, allowing an approximate model to absorb some
(although it can never be all) of the error caused by model misspecification
(in our case via an intrinsic dispersion). Naturally, the closer one's 
underlying model is to the truth, the smaller this probabilistic dispersion 
need be, and in the ultimate limit of a perfect model the probabilistic model 
tends towards a deterministic one. Since we do not make the claim that a broken 
power-law is the true description of the MR relation, the probabilistic model 
is essential for reliable forecasting, as it enables predictions in 
spite of the fact our model is understood to not represent the truth.

Whilst each of these three key features have been applied to MR relations
in some form independently, a novel quality of our methodology is to adopt
all three. For example, \citet{2015arXiv150407557W} inferred a probabilistic
power-law conditioned on the masses and radii of 90 exoplanets with radii below
$8$\,\rearth. This range crosses the expected divide between solid planets
and those with significant gaseous envelopes at $1.5$-$2.0$\,\rearth\ 
\citep{2014ApJ...792....1L} and so the authors tried truncating the data at 
$1.6$\,\rearth\ as an alternative model. In this work, we argue that the 
transitional points can actually be treated as free parameters in the model, 
enabling us to infer (rather than assume) their location and test theoretical 
predictions. Additionally, the data need not be censored at $<4$\,\rearth\ and 
the wider range makes a forecasting model less susceptible to truncation issues
at the extrema (we point out that \citet{2015arXiv150407557W} did not set out 
to develop a forecasting model explicitly, and thus this is not a criticism of 
their work, but rather just an example of how our work differs from previous 
studies).

\subsection{Data Selection}
\label{sub:dataselection}

Having broadly established the motivation (see Section~\ref{sec:intro}) and 
requirements (see Section~\ref{sub:philosophy}) for our model, we will use the rest 
of Section~\ref{sec:model} to provide a more detailed account of our methodology. To begin, 
we first define our basic criteria for a data point (a mass and radius 
measurement) to be included in what follows. Since our work focuses on the MR
relation, all included objects must fundamentally have a well-defined mass and 
radius. Whilst the former is universally true, the latter requires that the 
object have a nearly spherical shape. Low mass objects, for example the comet 
67P/Churyumov-Gerasimenko, may not have sufficient self-gravity to overcome
rigid body forces and assume a hydrostatic equilibrium shape (i.e. nearly 
spherical). The corresponding threshold mass limit should lie somewhere between
the most massive body which is known to not be in hydrostatic equilibrium
(Iapetus; $1.8\times10^{21}$\,kg; \citealt{GPMP}) and the least massive body 
confirmed to be in hydrostatic equilibrium (Rhea; $2.3\times10^{21}$\,kg; 
\citealt{GPMP}). This leads us to adopt a boundary condition of 
$M>2\times10^{21}$\,kg for all objects considered in this work.

As for the upper limit, we choose the maximum mass to be that of a star that
must still lie on the main-sequence within a Hubble time. The lifetime of a 
star is dependent upon its mass and luminosity, to first-order.
Given that the Sun will spend 10\,Gyr on the main-sequence and 
$L\propto M^{7/2}$, then the lifetime, $\tau \simeq (M/M_{\odot})^{-5/2} 
10\,\mathrm{Gyr} $.  This results in an upper limit of $M<0.87$\,\msun\
($1.7\times10^{30}$\,kg) for $\tau=H_0^{-1}$\,Gyr (where we set 
$H_0 = 69.7$\,km/s; \citealt{2014A&A...571A...1P}). Therefore, between our
lower and upper limits, there is a difference of nine orders-of-magnitude in 
mass and three order-of-magnitude in radius.

We performed a literature search for all objects within this range with a mass
and radius measurement available. For Solar System moons, we used
\href{http://www.dtm.ciw.edu/users/sheppard/satellites/}{The Giant Planet 
Satellite and Moon Page} \citep{GPMP} which is curated by Scott Sheppard
\citep{2003Natur.423..261S,2005AJ....129..518S,2006AJ....132..171S} and for the
planets we used the
\href{http://nssdc.gsfc.nasa.gov/planetary/factsheet/planet_table_ratio.html}{
NASA Planetary Fact Sheet} \citep{NPFS}. For extrasolar planets, we used the 
\href{http://nssdc.gsfc.nasa.gov/planetary/factsheet/planet_table_ratio.html}{
TEPCat} catalog of ``well-studied transiting planets'', curated by John 
Southworth \citep{2008MNRAS.386.1644S,2009MNRAS.394..272S,2010MNRAS.408.1689S,
2011MNRAS.417.2166S,2012MNRAS.426.1291S}. Brown dwarfs and low-mass stars were
drawn from a variety of sources, which we list (along with all other objects
used in this work) in \mytable\ref{tab:data}.

In order to later fit these data sources to an MR model, it is necessary to
define a likelihood function of each datum. We later (see 
\S\ref{sub:likelihood}) make the assumption that for a quoted mass (or radius)
measurement of $M = (a \pm b)$, that one reasonably approximate
$M\sim\mathcal{N}(a,b)$. This assumption is a poor one for low signal-to-noise
data, especially for upper limit constraints only, where $M$ (or $R$) is more
likely to follow an asymmetric profile centered near zero. Without knowledge of
the correct likelihood function, we argue that such data are best excluded in
what follows.

For this reason, we apply a 3\,$\sigma$ cut to both mass ($(\omob/\omerr)>3$)
and radius ($(\orob/\orerr)>3$). In what follows, we assume that both the mass
and radius measurements follow a normal distribution, which is symmetric. For
those data which have substantially asymmetric errors ($\errup \neq \errdown$)
then, we only use cases where the errors differ by less than 10\% (i.e.
$(| \errup - \errdown |)/(\errmean) \leq 0.1$). Together, these cuts 
remove 16\% of the initial data, which, as discussed later in \S\ref{sub:bias},
do not bias (or even noticeably influence) our final results.. Next, we take
the average of both errors $\errmean$ as the standard deviation of the normal 
distribution. In the end, we have \Ndata\ of objects in total which are listed in 
\mytable\ref{tab:data}.

The data span a diverse range of environments, with a variety orbital periods,
insolations, metallicities, etc. Since these terms are not used in our 
analysis, the results presented here should be thought of as a MR relation
marginalized over all of these other terms. Once again, we stress that the
effects of these terms is naturally absorbed by the probabilistic framework of 
our model, meaning that forecasts may be made about any new data, provided it
can be considered representative of the data used for our analysis.

\subsection{Probabilistic Broken Power-Law}
\label{sub:powerlaw}

We elect to model the MR relation with a probabilistic broken power-law, for
the reasons described in \S\ref{sub:philosophy}. By probabilistic, we mean that
this model includes intrinsic dispersion in the MR relation to account for
additional variance beyond that of the formal measurement uncertainties. This
dispersion represents the variance observed in nature itself around our broken
power-law model. To put this in context, a deterministic MR power-law would be
described via

\begin{equation}
\frac{\orob}{\re} =  \powerconst \Big(\frac{\omob}{\me}\Big)^\slope,
\label{power_deter}
\end{equation}

where $\orob$ \& $\omob$ are the mass and radius of the object respectively and
$\powerconst$ \& $\slope$ are the parameters describing the power-law. However, it is easy 
to conceive of two objects with the exact same mass but different compositions, thereby 
leading to different radii. For this reason, we argue that a deterministic model
provides a unrealistic description of the MR relation. In the probabilistic model, 
for any given mass there is a corresponding distribution of radii. In this work, we assume a
normal distribution in the logarithm of radius. The mean of the distribution takes the
result of the deterministic model, and the standard deviation is the intrinsic dispersion, 
a new free parameter.

A power-law relation can be converted to a linear relation by taking logarithm on both axes. 
In practice, we take the logarithm base ten of both mass and radius in Earth units, and use a
linear relation to fit them. In what follows, we will use $\omob$, $\orob$ to 
represent mass and radius, and $\logm$ and $\logr$ to represent $\logmeq$ and $\logreq$.
The power-law relation turns into

\begin{equation}
\logr = \offset + \logm \times \slope,
\label{linear_deter}
\end{equation}

where $\logr=\logreq$, $\logm=\logmeq$ and $\offset = \log_{10} \powerconst$.
In what follows, we will use $\normal(\mu,\sigma)$ as the normal distribution, where
$\mu$ is the mean and $\sigma$ is the standard deviation.
The corresponding probabilistic relation in $\log$ scale becomes

\begin{equation}
\logr \sim \normal (\mu = \offset+\logm \times \slope, \sigma = \scatter)
\label{linear_prob}
\end{equation}

On a logarithmic scale, the data still approximately follow normal distributions, 
because the logarithm of a normal distribution is approximately a normal distribution
when the standard deviation is small relative to the mean, which is true here
since we made a 3\,$\sigma$ cut in both mass and radius. The original data, 
$\omob \sim \normal ( \omt, \omerr)$, will turn into $\mob \sim \normal (\mt, \merr)$, 
where $\mt = \log_{10} (\omt / \me)$ and $\merr = \log_{10}(e) (\omerr/\omob)$.

We consider it more reasonable to assume that the intrinsic dispersion in radius
will be a fractional dispersion, rather than an absolute dispersion. For example, the
dispersion of Earth-radius planets might be $\mathcal{O}[0.1\,R_{\oplus}]$ but for
stars it should surely be much larger in an absolute sense. Since a fractional
dispersion on a linear scale corresponds to an absolute dispersion on logarithmic
scale, this assumption is naturally accounted for by our model. To implement
the probabilistic model, we employ a hierarchical Bayesian model, or HBM for short.

\subsection{Hierarchical Bayesian Modeling}
\label{sub:HBM}

\begin{figure}
\begin{center}
\includegraphics[width=8.4cm,angle=0,clip=true]{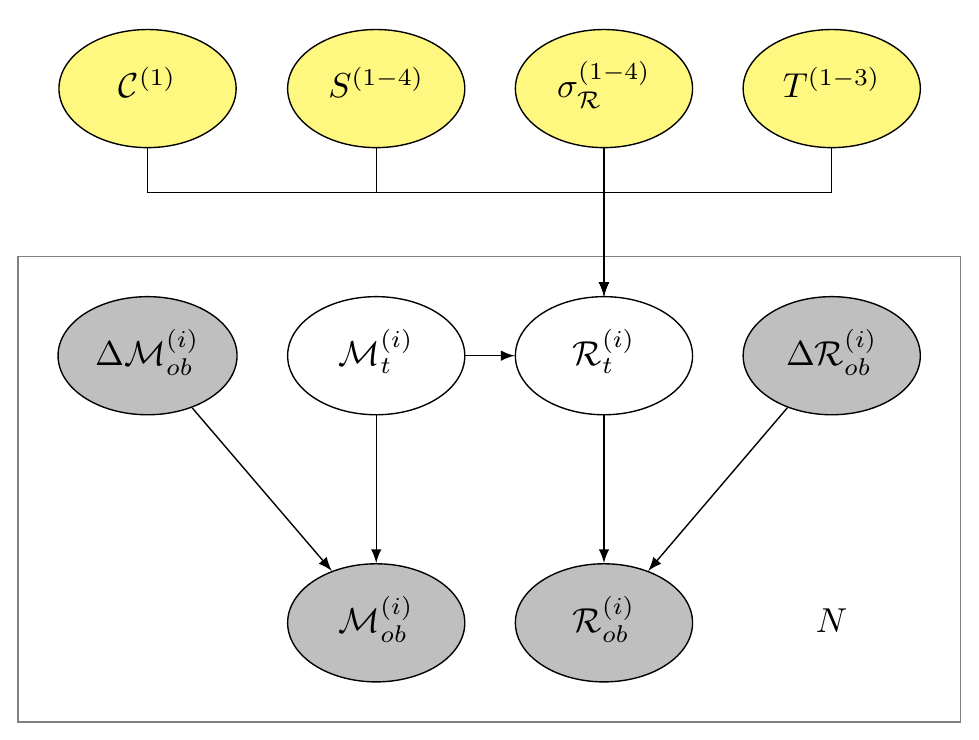}
\caption{
Graphical model of the HBM used to infer the probabilistic MR relation
in this work. Yellow ovals represent hyper-parameters, white represent the 
true local parameters and gray represent data inputs. All objects on
the plate have $N$ members.
}
\label{cartoon}
\end{center}
\end{figure}

The difference between an HBM and the more familiar Bayesian method is that HBMs
have two sets of parameters; a layer of hyper parameters, $\hyper$, on top of the
local parameters, $\local$ (see \citealt{2010ApJ...725.2166H} for a pedagogical
explanation). The local parameters usually describe the properties 
of each individual datum, whilst the hypers describe the overall ensemble properties.
For example, in this work, the local parameters are the true $\logmeq$, $\logreq$
(or $\mt$, $\rt$) of all the objects, and the hyper parameters, $\hyper$, are those
that represent the broken power-law. This hierarchical structure is illustrated 
in \myfig\ref{cartoon}, which may be compared to the analogous graphical model shown in Figure 1 of
\citet{2015arXiv150407557W}.

Some of the first applications of this method are \citet{1995ApJS...96..261L},
\citet{1996AIPC..366..196G}, and \citet{2010ApJ...725.2166H} (in exoplanets research). 

For the local parameters, we define a mass, $\mt$, and radius, $\rt$, term for
each object giving \Nxdata\ local variables. In practice, the $\rt$ local 
parameters are related to the $\mt$ term through the broken power-law and each
realization of the hyper parameters. In total then, our model includes \Nxdata\
local parameters and a compact set of hyper parameters, as described later in 
the MCMC subsection.

\subsection{Continuous Broken Power-Law Model}

Plotting the masses and radii on a $\log$-$\log$ scale, (as shown later in
Figure~\ref{fig:MR}), it is clear that single, continuous power-law is unable to
provide a reasonable description of the data. For example, one might reasonably
expect that the Neptune-like planets follow a different MR relation from the
terrestrial planets, since the voluminous gaseous envelope of the former 
dominates their radius \citep{2014ApJ...792....1L}. This therefore argues in 
favor of using a segmented (or broken) power-law.

At least three fundamentally distinct regimes are expected using some
simple physical insights; a segment for terrestrial planets, gas giants and 
stars. Indeed, the MR data clearly shows distinct changes in the power-index,
corresponding to the transition points between each segment. However, a visual
inspection also reveals a turn-over in the MR relation at around a Saturn-mass.
Therefore, from one roughly Saturn-mass to the onset of stars, there is a 
strong case for a fourth segment which we consequently include in our model. 
Later, in Section~\ref{sub:modelcomp}, we perform a model comparison of
a three- versus four-segment model to validate that the four-segment broken 
power-law is strongly favored.

Our favored model consists of 12 free hyper parameters; 1 offset ($\offset^(1)$),
4 slopes ($\slope \onetofour$), 4 intrinsic dispersions ($\scatter \onetofour$),
and 3 transition points ($\transition \onetothree$). Critically then, we
actually fit for the location of transition points and include an independent
intrinsic dispersion for each segment (making our model probabilistic). Also
note that the ``slopes'' in $\log$-$\log$ space are the power-law indices in
linear space. The hyper parameter vector is therefore

\begin{align}
\hyper = \{&\slope^{(1)}, \slope^{(2)}, \slope^{(3)}, \slope^{(4)},\nonumber\\
\qquad& \scatter^{(1)}, \scatter^{(2)}, \scatter^{(3)}, \scatter^{(4)},\nonumber\\
\qquad& \transition^{(1)},\transition^{(2)},\transition^{(3)},\offset^{(1)}\}.
\label{hyper_set}
\end{align}

There is only one free parameter for the offset since we impose the condition
that each segment of the power-law is connected, i.e. a continuous broken power-law. 
By requiring that two segments meet at the transition point between them, we can
derive the offsets for the rest of the segments. At each transition point
$\transition \jth$,

\begin{equation}
\offset \jth + \slope \jth \times \transition \jth = \offset \kth + \slope \kth \times \transition \jth
\quad \rm{for}\ j = 1,2,3.
\label{derive_c}
\end{equation}

We can now iteratively derive the other offsets as,

\begin {equation}
\offset \kth = \offset \jth + (\slope \jth - \slope \kth) \times \transition \jth
\quad \rm{for }\ j = 1,2,3.
\label{express_c}
\end {equation}

\subsection{Hyper Priors}

The hyper priors, that is the priors on the hyper-parameters, are selected
to be sufficiently broad to allow an extensive exploration of parameter space
and to be identical for each segment. Uniform priors are used for the location
parameters, namely the offset, $\offset$, and transition points, $\transition$.
For scale parameters, namely the intrinsic dispersion $\scatter$, we adopt
log-uniform priors.

For the slope parameters, we don't want to constrain them in a specific range, so
we use the normal distribution with a large variance. This leads to a prior which is
approximately uniform in any small region yet loosely constrains the MCMC walkers to
the relevant scale of the data. A detailed list of the priors is provided in 
\mytable \ref{hyper_table}.
 
\subsection{Two Different Categories of Local Parameters}

The local parameters in our model are formally $\mt$ and $\rt$, 
although in practice $\rt$ doesn't need to be fitted explicitly since
it is derived from the realization of the broken power-law (as described in more detail later).

Even for $\mt$ though, there are two categories that we must distinguish between.
Objects within the Solar System tend to have very precise measurements of their
fundamental properties such that their formal uncertainties are negligible
relative to the uncertainties encountered for extrasolar objects, for which we
must account for the measurement uncertainty in our model.

For objects with negligible error, we simply fix $\mt = \mob$ and
$\rt = \rob$, since $\merr, \rerr \propto \frac{\omerr}{\omob}, 
\frac{\orerr}{\orob} = 0$. For objects in the second category, $\mt$ are 
set to be independently uniformly distributed in $[-4, 6]$. Throughout the
paper, we will use $\uniform(a,b)$ to denote a uniform distribution, where
$a$ and $b$ are the lower and upper bounds of the distribution, so for
example

\begin{equation}
\mt \ith \sim \uniform(-4,6)
\quad \rm{for}\ i = 1,2,...,\Ndata.
\label{mt_prior}
\end{equation}

\subsection{Inverse Sampling}

We use the inverse sampling method to sample the parameters $\mt$ and $\hyper$.
By inverse sampling, we mean that the walkers directly sample in the 
probability space, rather than the parameter space itself. By directly
walking in the prior probability space with Gaussian function as our proposal distribution, 
inverse sampling is more efficient
than walking in real space plus likelihood penalization (see \citealt{inverse}
further details on the inverse sampling method).

For each jump in the MCMC chain, we sample a probability, $\prob$, for each
parameter with $\uniform(0,1)$. 
We then determine this parameter's cumulative distribution from its prior
probability distribution. With $\prob$ and the cumulative distribution, we can
then calculate the corresponding sample of the parameter.

The equations of the prior distributions of $\mt$ and $\hyper$ are already 
shown in \mytable \ref{hyper_table} and \myeqn(\ref{mt_prior}). With inverse
sampling, the effects of the priors have already been accounted for, meaning that
we do not need to add the prior probabilities of a parameter into the total
log-likelihood function.

\subsection{Total Log-Likelihood}
\label{sub:likelihood}

As discussed above, since $\mt$ and $\hyper$ are drawn with inverse sampling,
then there is no need to add corresponding penalty terms to the log-likelihood
function. The total log-likelihood is now based on how we sample $\rt$ from $\mt$ 
and $\hyper$, and the relations between $\mt$, $\rt$ and data. The relations are 
given by

\begin{equation}
\mob \ith \sim \normal \Big(\mt \ith, \merr \ith \Big).
\label{mob_dist}
\end{equation}

When $\merr \ith=0$, the above equation can be interpreted as 
$\mob \ith = \mt \ith$, which corresponds to the case where measurement 
errors are zero. This is also true for $\rob \ith$, such that

\begin{equation}
\rob \ith \sim \normal \Big(\rt \ith, \rerr \ith \Big),
\label{rob_dist}
\end{equation}

and

\begin{equation}
\rt \ith \sim \normal \Big( f(\mt \ith,\hyper), \scatter' \Big),
\label{rt_dist}
\end{equation}

where we define

\begin{align}
\big(f(\mt \ith,& \hyper), \scatter' \big)=\nonumber\\
\qquad& \left\{ \begin{array}{ll}
\big(\offset \fst + \mt \ith \slope \fst, \scatter \fst \big) & \mt \ith \le \transition \fst \\
\big(\offset \snd + \mt \ith \slope \snd, \scatter \snd \big) &\transition \fst < \mt \ith \le \transition \snd \\
\big(\offset \trd + \mt \ith \slope \trd, \scatter \trd \big) &\transition \snd < \mt \ith \le \transition \trd \\
\big(\offset \fth + \mt \ith \slope \fth, \scatter \fth \big)&\transition \trd < \mt \ith  \end{array}
       \right. .\label{fh_case}
\end{align}

Combining \myeqn(\ref{rob_dist}) and (\ref{rt_dist}), we have

\begin{equation}
\rob \ith \sim \normal \Big( f(\mt \ith, \hyper), \sqrt{ (\rerr \ith)^2 + (\scatter')^2} \Big).
\label{rob_combine}
\end{equation}

\myeqn(\ref{rob_combine}) shows that if we have already sampled $\mt$ and $\hyper$, we don't need to sample $\rt$ anymore since $\rob$ can be directly related to $\mt$ and $\hyper$.
From \myeqn(\ref{mob_dist}) and (\ref{rob_combine}), we can see that the total log-likelihood of the model is

\begin{align}
-2\log \mathcal{L} =& \sum_{i=1}^N \Bigg( \frac{\mob \ith - \mt \ith}{\merr \ith} \Bigg)^2 + \sum_{i=1}^N (\merr \ith)^2\\
\qquad& \sum_{i=1}^N \frac{ \Big( \rob \ith - f(\mt \ith, \hyper) \Big)^2}{ \Big( \rerr \ith \Big)^2 + \Big( \scatter' \Big)^2} + \nonumber\\
\qquad& \sum_{i=1}^N \log \Big[ \big( \rerr \ith \big)^2 + \big( \scatter' \big)^2 \Big].
\label{total_loglike}
\end{align}

Note that in the above, we assume mass and radius have no covariance, which is almost
always true given the independent methods of their measurement.

\section{ANALYSIS}
\label{sec:analysis}

\subsection{Parameter Inference with Markov Chain Monte Carlo}

We used the Markov Chain Monte Carlo (MCMC) method with the Metropolis 
algorithm \citep{1953JChPh..21.1087M} to explore the parameter space and infer 
the posterior distributions for both the hyper and local parameters. The 
Metropolis algorithm uses jumping walkers, proceeding by accepting or rejecting
each jump by comparing its likelihood with that of the previous step. Since we 
have 12 hyper parameters and \Ndata\ data points (corresponding to \Ndata\ 
$\mt$), the walker jumps in a probability hyper cube of (12+\Ndata) dimensions.

We begin by running \ninitial\ independent initial chains for 500,000 accepted
steps each, seeding the parameters $\transition\onetothree$ from 0.5, 2, and 4
with Gaussian distributions of sigma one (but keeping all others
terms drawn seeded from a random sample from the hyper priors).

We identify the burn-in point by eye, searching for the instant where the local
variance in the log-likelihood (with respect to chain step) stabilizes to a 
relatively small scatter in comparison to the initial steps. This burn-in
point tended to occur after $\simeq200,000$ accepted steps, largely driven by the fact
that both the hyper and local parameters were not seeded from a local minimum
(with the exception of $\transition\onetothree$) and therefore required
a substantial number of steps to converge.

Combining these initial chains, we chose \ncontinue\ different realizations 
which have the highest log-likelihood but also not too close to each other.
We then start \ncontinue\ new independent chains, where each chain is seeded 
from one of the top 200 log-like solutions found from the stacked initial 
chains. We run each of these \ncontinue\ chains for $10^7$ trials
with acceptance rate $\sim 5\%$ (i.e. 500,000 accepted steps each)
and find, as expected, that each chain is burnt-in right from the beginning.

\begin{figure*}
\begin{center}
\includegraphics[width=18.0cm,angle=0,clip=true]{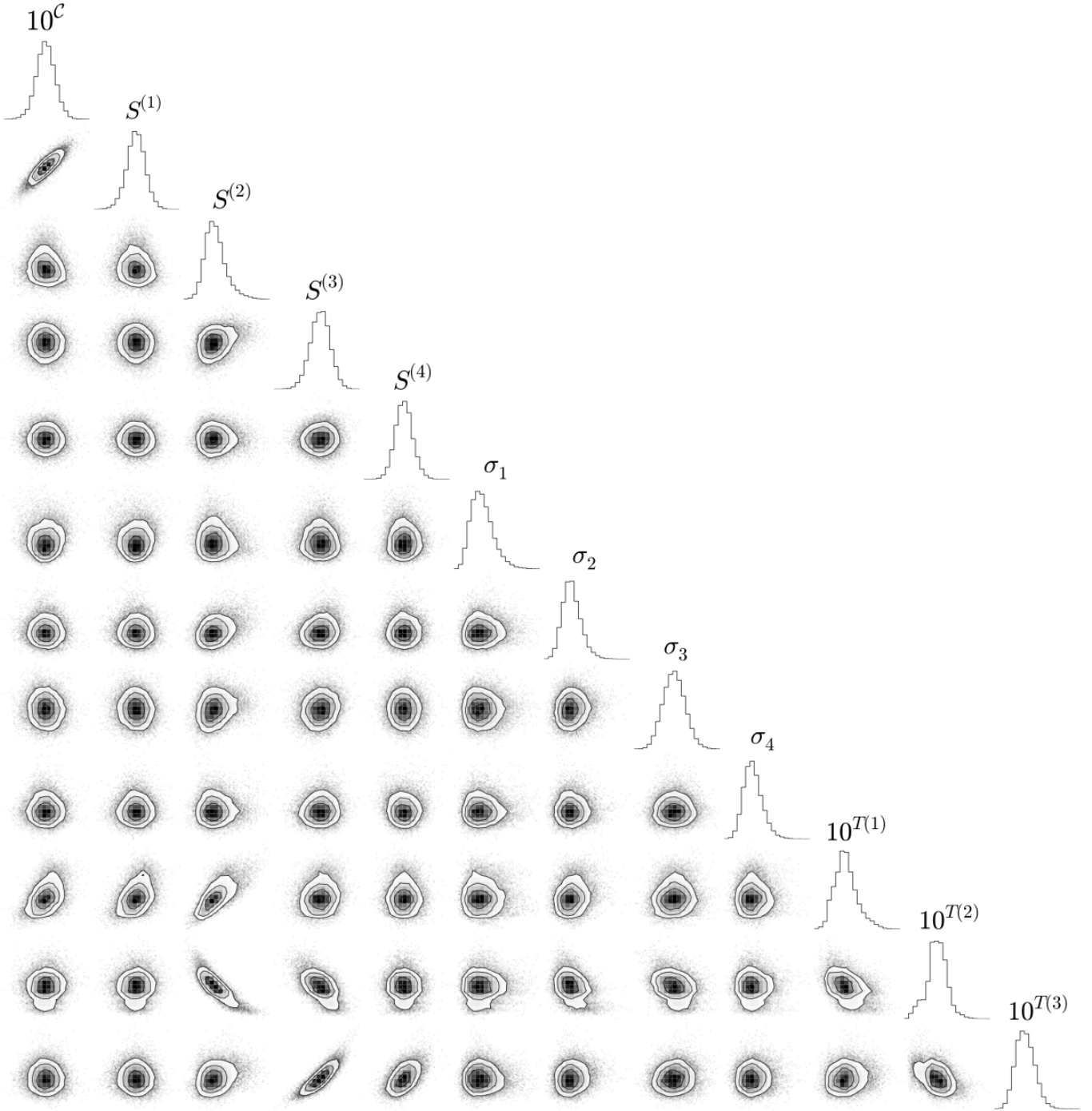}
\caption{
Triangle plot of the hyper-parameter joint posterior distribution
(generated using \href{https://github.com/dfm/corner.py.git}{corner.py}). Contours
denote the 0.5, 1.0, 1.5 and 2.0\,$\sigma$ confidence intervals.
}
\label{fig:triangle}
\end{center}
\end{figure*}

\begin{figure*}
\begin{center}
\includegraphics[width=18.0cm,angle=0,clip=true]{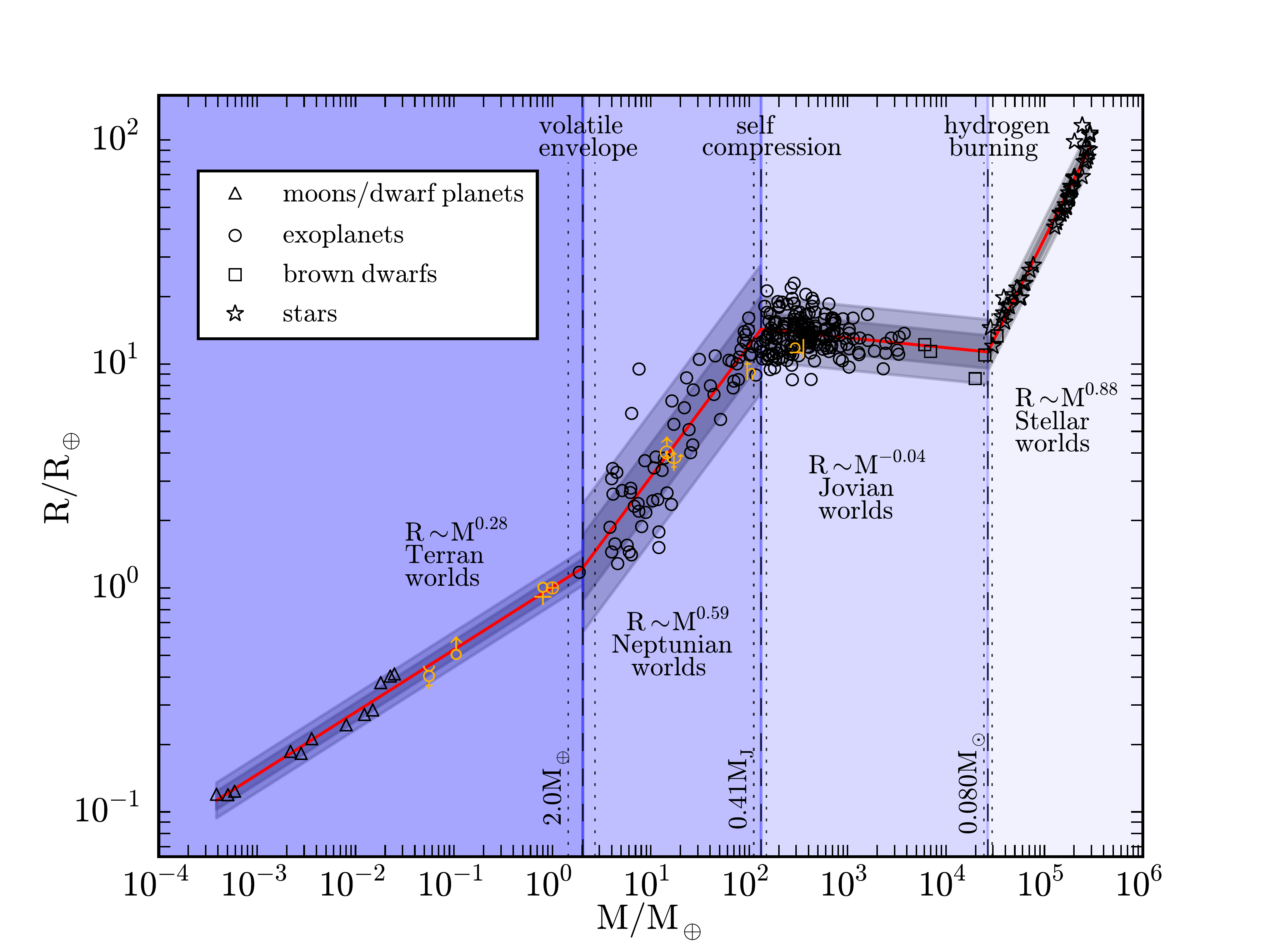}
\caption{
The mass-radius relation from dwarf planets to late-type stars. Points 
represent the \Ndata\ data against which our model is conditioned, with the
data key in the top-left. Although we do not plot the error bars, both radius
and mass uncertainties are accounted for. The red line shows the mean
of our probabilistic model and the surrounding light and dark gray regions 
represent the associated 68\% and 95\% confidence intervals, respectively. The
plotted model corresponds to the spatial median of our hyper parameter 
posterior samples.
}
\label{fig:MR}
\end{center}
\end{figure*}

To check for adequate mixing, we calculated the effective length, defined as
the length of the chain divided by the correlation length, where the 
correlation length is defined as $\lcorr = \min_{lag} \{ | 
\mathrm{AutoCorrelation}(\mathrm{chain}, \mathrm{lag}) | < 0.5 \}$ 
\citep{2004PhRvD..69j3501T}. We find that the sum of the effective lengths 
exceeds 2000 (i.e. is $\gg1$), indicating good mixing. We also verified
that the Gelman-Rubin statistic \citep{Gelman92} dropped below 1.1 (it was 1.02), 
indicating that the chains had converged. Finally, we thinned 
the \ncontinue\ chains by a factor of 100, and stacked them together, which 
gives a combined chain of length of $10^6$. The hyper-parameter posteriors, 
available at \href{\packagelink}{this URL}, are shown as a triangle plot in 
Figure~\ref{fig:triangle}. We list the median and corresponding 68.3\% 
confidence interval of each hyper parameter posterior in 
\mytable~\ref{hyper_table}. Our model, evaluated at the spatial median of the
hyper parameters, is shown in Figure~\ref{fig:MR} compared to the data upon 
which it was conditioned. The spatial median simply finds the sample from
the joint posterior which minimizes the Euclidean distance to all other
samples.

\begin{deluxetable*}{cll}
\tablecolumns{8}
\tablewidth{0pc}
\tablecaption{
Description and posterior of hyper parameters. 
The prior distributions of the hyper parameters are 
$\offset^{(1)} \sim \uniform (-1,1)$;
$\slope \onetofour \sim \normal (0,5)$;
$\log_{10} \big[\scatter \onetofour\big] \sim \uniform (-3,2)$; 
$\transition \onetothree \sim \uniform (-4,6)$.
}
\tablehead{
\colhead{$\hyper$ term} & \colhead{Description}   & \colhead{Credible Interval}
}
\startdata
$10^{\offset}$\,\rearth& Power-law constant for the \terr\ (T-class) worlds MR relation		& $1.008_{-0.045}^{+0.046}$\,\rearth \\
\hline
$\slope \fst$      & Power-law index of \terr\ worlds; $R \propto M^{\slope}$ & $0.2790_{-0.0094}^{+0.0092}$  \\ 
$\slope \snd$      & Power-law index of \nept\ worlds; $R \propto M^{\slope}$ & $0.589_{-0.031}^{+0.044}$  \\
$\slope \trd$      & Power-law index of \jovi\ worlds; $R \propto M^{\slope}$ & $-0.044_{-0.019}^{+0.017}$ \\
$\slope \fth$      & Power-law index of \stel\ worlds; $R \propto M^{\slope}$ & $0.881_{-0.024}^{+0.025}$  \\
\hline
$\scatter \fst$    & Fractional dispersion of radius for the \terr\ MR relation & $4.03_{-0.64}^{+0.94}$\,\% \\
$\scatter \snd$    & Fractional dispersion of radius for the \nept\ MR relation & $14.6_{-1.3}^{+1.7}$\,\%  \\
$\scatter \trd$    & Fractional dispersion of radius for the \jovi\ MR relation & $7.37_{-0.45}^{+0.46}$\,\% \\
$\scatter \fth$    & Fractional dispersion of radius for the \stel\ MR relation & $4.43_{-0.47}^{+0.64}$\,\%  \\
\hline
$10^{T(1)}$\,\mearth & \terr-to-\nept\ transition point & $2.04_{-0.59}^{+0.66}$ \mearth \\
$10^{T(2)}$\,\mearth & \nept-to-\jovi\ class transition point & $0.414_{-0.065}^{+0.057}$ \mjup \\
$10^{T(3)}$\,\mearth & \jovi-to-\stel\ class transition point & $0.0800_{-0.0072}^{+0.0081}$ \msun
\enddata
\label{hyper_table}
\end{deluxetable*}

\subsection{Model Comparison}
\label{sub:modelcomp}

The model with four segments was at first selected by visual inspection of the
data. Two of the three transition points, $T(1)$ and $T(3)$, occur at locations
which can be associated with physically well-motivated boundaries
(planets accreting volatile envelopes, \citealt{2015ApJ...801...41R}, and hydrogen
burning, \citealt{2014AJ....147...94D}), whereas the $T(2)$ transition
is not as physically intuitive.

In order to demonstrate that this model is statistically favored over the
three-segment model, we repeated all of the fits for a simpler three-segment 
model. We seed the remaining two transition points from the approximate 
locations of $T(1)$ and $T(3)$ found from the four-segment model fit. We label 
this model as $\mathcal{H}_3$ and the four-segment model used earlier as 
$\mathcal{H}_4$.

For this simpler model, $\mathcal{H}_3$ uses only two transition points which 
breaks the data into three different segments. To implement this model, the 
only difference is that the hyper parameters vector $\hyper'$ becomes

\begin{equation}
\hyper' = (\offset, \slope \onetothree, \scatter \onetothree, \transition \onetotwo).
\end{equation}

We find that the maximum log-likelihood of $\mathcal{H}_3$ is considerably less
than that of $\mathcal{H}_4$, less by \logdiff\ (corresponding to $\Delta\chi^2
= 68.3$ for \Ndata\ data points) at the gain of just three fewer free 
parameters. The marginal likelihood cannot be easily computed in the very 
high dimensional parameter space of our problem, and the Bayesian and Akaike 
information criteria (BIC, \citealt{BIC} \& AIC, \citealt{AIC}) are also
both invalid for such high dimensionality. Instead, we used the Deviance 
information criterion (DIC, \citealt{DIC}), a hierarchical modeling 
generalization of the AIC, to compare the two models. When comparing two
models with the DIC, the smaller value is understood to the preferred model.
We find that $\mathrm{DIC}(\mathcal{H}_4)=-665.5$ and
$\mathrm{DIC}(\mathcal{H}_3)=-333.5$, indicating a strong preference for model
$\mathcal{H}_4$.

\subsection{
The Effect of our Data Cuts
}
\label{sub:bias}

As discussed earlier in \S\ref{sub:dataselection}, our data cuts removed
16\% of the initial data considered. Since these points are low SNR data,
they, by definition, have a weak effect on the likelihood function. As
evident in Figure~\ref{fig:MR}, there is an abundance of precise data
constraining the slope parameters in each segment and none of the segments
can be described as residing in a poorly constrained region. Given that the
transition points are defined as the intercept of the slopes, they too
are well constrained by virtue of the construction of our model. Critically,
then, a paucity of data at the actual transition point locations (as is true 
for $T(1)$) has little influence on our inference of their locations. 
In order for the results of this work to be significantly affected by the 
exclusion of these low SNR data then, these points would have to have modify 
the inference of the slope parameters.

To demonstrate this effect is negligible, we consider the Neptunian 
segment in isolation, since it strongly affects the critical transition 
$T(1)$ and features the largest fraction of excluded points (24\%).
Since the excluded data were due to lossy mass measurements, we ignore
the radius errors and perform a simple weighted linear least squares
regression with and without the excluded data, where we approximate
the observations to be normally distributed. We find that
the slope parameter, $S(2)$, changes from $0.782\pm0.058$ to 
$0.784\pm0.050$ by re-introducing the excluded data, illustrating
the negligible impact of these data.

\subsection{
Injection/Recovery Tests
}
\label{sub:inject}

In order to verify the robustness of our algorithm, we created
ten fake data sets and blindly ran our algorithm again on each. The
data sets are generated by making a random, fair draw from our
joint posteriors (both the local and hyper parameters), ensuring
that each draw is from an different effective chain. We then re-ran
our original algorithm as before, except the number of steps in the
final chain is reduced by a factor of ten for computational expedience.

\begin{figure*}
\begin{center}
\includegraphics[width=17.4cm,angle=0,clip=true]{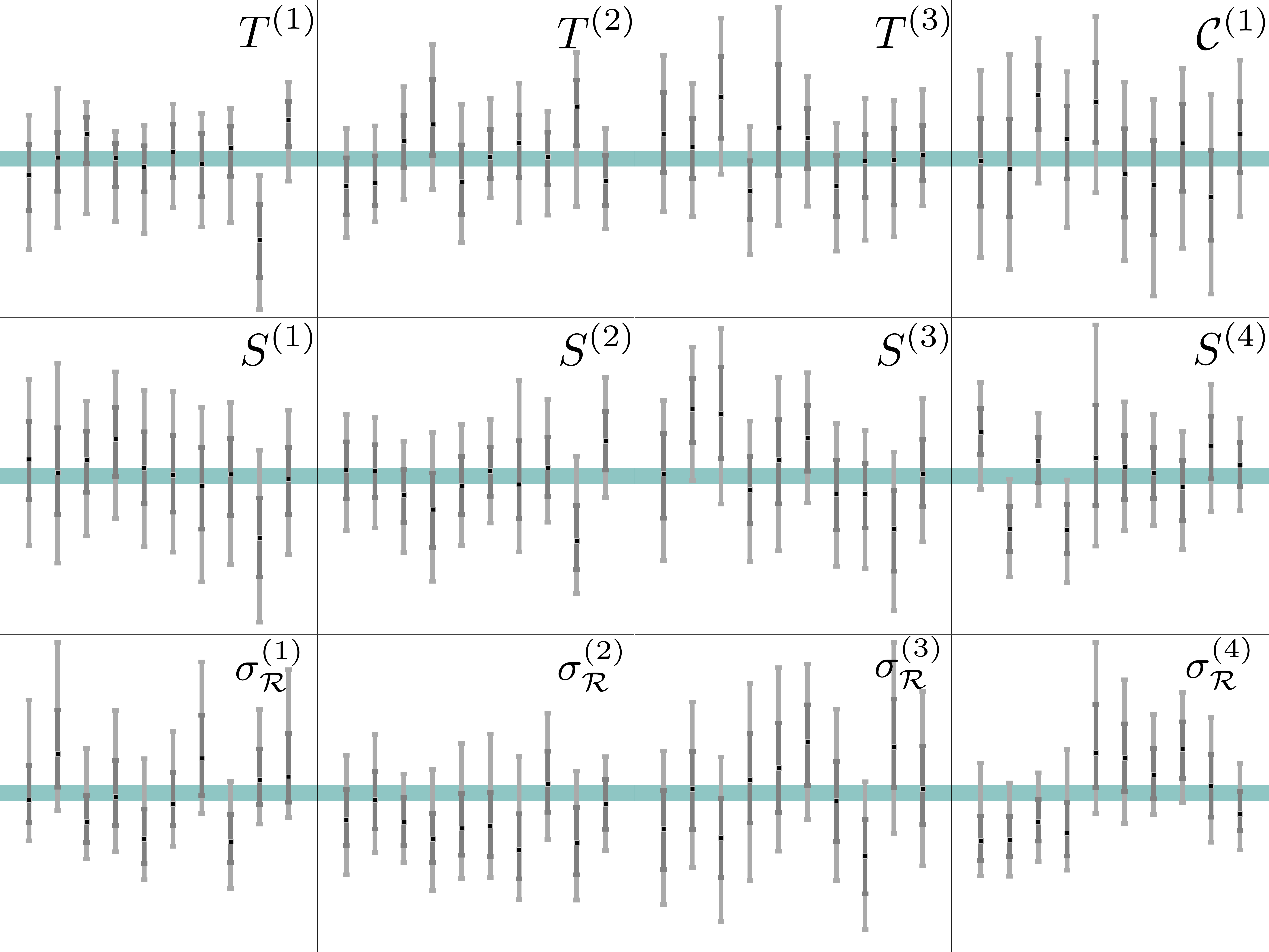}
\caption{
Each sub-panel shows the residuals of a hyper-parameter in our model, as 
computed between ten injected truths and the corresponding recovered values. 
The black square denotes the recovered posterior median and the dark \& light
gray bars denote the 1 \& 2\,$\sigma$ confidence intervals. The green
horizontal bar marks the zero-point expected for a perfect recovery.
}
\label{fig:inject}
\end{center}
\end{figure*}

We computed the one and two-sigma credible intervals on each 
hyper-parameter and compare them to the injected truth in 
Figure~\ref{fig:inject}. As evident from this figure, we are able to
easily recover all of the inputs to within the expected range,
validating the robustness of the main results presented in this work.

\section{CLASSIFICATION}
\label{sec:classification}

\subsection{Classification with an MR relation}

A unique aspect of this work was to use freely fitted transitional points in 
our MR relation. As discussed earlier, these transitional points essentially 
classify the data between distinct categories, where the class boundaries 
occur in mass and are defined using the feature of 
$\mathrm{d}\mathcal{R}/\mathrm{d}\mathcal{M}$. Such classes are evident even 
from visual inspection of the MR data (see Figure~\ref{fig:MR}), but our 
Bayesian inference of a self-consistent probabilistic broken power-law provides
statistically rigorous estimates of these class boundaries. In what follows, we
discuss the implications of the inferred locations of the class boudnaries 
($T^{(\mathrm{1})}$, $T^{(\mathrm{2})}$ and $T^{(\mathrm{3})}$).

\subsection{Naming the Classes}

Rather than refer to each class as segments 1, 2, 3 and 4, we here define a
name for each class to facilitate a more physically intuitive discussion of
the observed properties. A naming scheme based on the physical processes
operating is appealing but ultimately disingenuous since our model is
deliberately chosen to be a data-driven inference, free of physical assumptions
about the mechanics and evolution sculpting these worlds. We consider it
more appropriate, then, to name each class based upon a typical and well-known 
member.

For segment 2, Neptune and Uranus are typical members and are of course very
similar to one another in basic properties. We therefore consider this class
to define a sub-sample of Neptune-like worlds, or ``\nept'' worlds more 
succinctly. Similarly, we identify Jupiter as a typical member of segment 3, 
unlike Saturn which lies close to a transitional point. Accordingly, we define 
this sub-sample to be representative of Jupiter-like worlds, or ``\jovi'' 
worlds.

For the hydrogen-burning late-type stars of segment 4, these objects can be 
already classified by their spectral types spanning M, K and late-type G
dwarfs. Rather than refer to them as M/K/late-G class stars, we simply label 
them as stars for the sake of this work and for consistency with the 
``worlds'' taxonomy dub them ``\stel'' worlds.

Finally, we turn to segment 1 which is comprised largely of Solar System
members and thus all of which are relatively well-known. The objects span
dwarf planets to the terrestrial planets, silicate worlds to icy worlds,
making naming this broad class quite challenging. Additionally, calling this
class Earth-like worlds would be confusing given the usual association of
this phrase with habitable, Earth-analogs. For consistency with
the naming scheme used thus far, we decided that dubbing these objects as
``\terr'' worlds to be the most appropriate.

\subsection{$T^{(\mathrm{1})}$: The \Terr-\Nept\ Worlds Divide}
\label{sub:T1}

From masses of $\sim10^{-4}$\,\mearth\ to a couple of Earth
masses, we find that a continuous power-law of $R \sim M^{0.279\pm0.009}$
provides an excellent description of these \terr\ worlds. No break is
observed between ``dwarf planets'' and ``planets''. If the \terrs\
displayed a constant mean density, then we would expect $R\sim M^{1/3}$, and
so the slightly depressed measured index indicates modest compression with 
increasing mass ($\rho \sim M^{0.16\pm0.03}$). Our result is in close agreement
with theoretical models, which typically predict $R \sim M^{0.27}$ for 
Earth-like compositions (e.g. see \citealt{2006Icar..181..545V}). 

We find the first transition to be located at $(2.0\pm0.7)$\,\mearth,
defining the transition from \terrs\ to \nepts.
After this point, the density trend reverses with $\bar{\rho}\sim
M^{-0.77\pm0.13}$, indicating the accretion of substantial volatile
gas envelopes. This transition is not only evident in the power-law index,
but also in the intrinsic dispersion, which increases by a factor of
$(3.6\pm0.9)$ from \terrs\ to \nepts. This transition point is of major
interest to the community, since it caps the possibilities of rocky, habitable
Super-Earth planets, with implications for future missions designs
(e.g. see \citealt{2015arXiv150704779D}).

Our result is compatible with independent empirical and theoretical
estimates of this transition. Starting with the former, we compare our
result to \citet{2015ApJ...801...41R}, who sought the transition in
radius rather than mass. This was achieved by identifying radii which
exceed that of a solid planet, utilizing a principle first proposed 
by \citet{2013MNRAS.434.1883K}. Assuming an Earth-like compositional model, the
radius threshold was inferred to be $1.48_{-0.04}^{+0.08}$\,\rearth\
\citep{2015ApJ...801...41R}. Our result may be converted to a radius by using 
our derived relation. However, since our model imposes intrinsic radius 
dispersion (i.e. the probabilistic nature of our model), then the uncertainty 
in radius is somewhat inflated by this process. Nevertheless, we may convert
our mass posterior samples to fair radii realizations using our
\href{\packagelink}{\packagename} public code (described later in Section~\ref{sec:forecasting}). Accordingly,
we find that the transition occurs at $1.23_{-0.22}^{+0.44}$\,\rearth, which
is fully compatible with \citet{2015ApJ...801...41R}.

A comparison to theory comes from \citet{2014ApJ...792....1L}, who scale down
compositional models of gaseous planets to investigate the minimum size of a
H/He rich sub-Neptune. From this theoretical exercise, the authors estimate
that $1.5$\,\rearth\ is the minimum radius of a H/He rich sub-Neptune, which
is also compatible with our measurement. Therefore, despite the fact we 
do not impose any physical model (unlike \citealt{2014ApJ...792....1L}
\& \citealt{2015ApJ...801...41R}), our broken power-law model recovers the
transition from \terrs\ and \nepts.

\subsection{$T^{(\mathrm{3})}$: The \Jovi-\Stel\ Worlds Divide}
\label{sub:T3}

Another well-understood transition is recovered by our model at 
$(0.080\pm0.008)$\,\msun, which we interpret as the onset of hydrogen burning. 
As with the \terr-\nept\ worlds transition, we may compare this to other
estimates of the critical boundary. In the recent work of 
\citet{2014AJ....147...94D}, the authors performed a detailed observational 
campaign around this boundary. Inspecting the \Teff-$R$ plane, the authors 
identify a minimum at $\simeq0.086$\,\rsun, which corresponds to
$\simeq0.072$\,\msun with the 5\,Gyr isochrones\footnote{Although the point 
slightly precedes the first point in the \citealt{1998A&A...337..403B} grid, 
requiring a small linear extrapolation to compute.} of 
\citet{1998A&A...337..403B}. Based on this, we conclude that the result is 
fully compatible with our own prediction.

From stellar modeling, estimates of the minimum mass for hydrogen-burning 
range from 0.07\,\msun\ to 0.09\,\msun\ \citep{1993ApJ...406..158B,
1997ApJ...491..856B,1998A&A...337..403B,2000ApJ...542..464C,
2003A&A...402..701B,2008ApJ...689.1327S}. Therefore, both independent 
observational studies and theoretical estimates are consistent with our 
broken power-law estimate.

\subsection{$T^{(\mathrm{2})}$: The \Nept-\Jovi\ Worlds Divide}
\label{sub:T2}

We find strong evidence for a transition in our broken power-law at
$(0.41\pm0.07)$\,\mjup, corresponding to the transition between \nepts\
and \jovis. Whilst this transition has been treated as an assumed, fixed point
in previous works (e.g. \citealt{2013ApJ...768...14W} adopt a fixed transition
at 150\,\mearth, or 0.47\,\mjup), our work appears to be first instance
of a data-driven inference of this transition.

A plausible physical interpretation of this boundary is that \nepts\
rapidly grow in radius as more mass is added, depositing more gaseous
envelope to their outer layers. Eventually, the object's mass is sufficient
for gravitational self-compression to start reversing the growth,
leading into the \jovis. The existence of such a transition is not
unexpected, but our model allows for an actual measurement of its
location.

We infer the significance of this transition to be high at nearly 10\,$\sigma$
(see \S\ref{sub:modelcomp}), motivating us to propose that this transition is 
physically real and that a class of \jovis\ is taxonomically 
rigorous in the mass-radius plane. a defining feature of the \jovi\ worlds
is that the MR power-index is close to zero ($-0.04\pm0.02$), with radius being
nearly degenerate with respect to mass.

We find that brown dwarfs are absorbed into this class, displaying
no obvious transition (also see Figure~1) at $\sim13$\,\mjup, the canonical
threshold for brown dwarfs \cite{2011ApJ...727...57S}, as was also argued by
\cite{2015arXiv150605097H}. When viewed in terms of mass and radius then, brown
dwarfs are merely high-mass members of a continuum of \jovis\ and more closely
resemble ``planets'' than ``stars''.

The fact that the \nept-to-\jovi\ transition occurs at around one
Saturn mass is generally incompatible with theoretical predictions of
a H/He rich planet, such as Saturn. Calculations by \cite{1969ApJ...158..809Z}
predict that a cold sphere of H/He is expected to reach a maximum size
somewhere between $1.2$\,\mjup\ to $3.3$\,\mjup. The suite of models
produced by \cite{2007ApJ...659.1661F} for H/He rich giant planets, for various
insolations and metallicities, peak at masses between $\sim2.5$\,\mjup\
to $\sim5.2$\,\mjup. Nevertheless, Jupiter and Saturn have similar
radii (within 20\%) of one another despite a factor of three difference
in mass, crudely indicating that \jovis\ commence at a mass less than
or equal to that of Saturn.

\section{FORECASTING}
\label{sec:forecasting}

\subsection{\href{\packagelink}{\packagename}: An Open-Source Package}

Using our probabilistic model for MR relation inferred in this work, it is 
possible to now achieve our primary objective: to forecast the mass (or 
radius) of an object given the radius (or mass). Crucially, our
forecasting model can not only propagate measurement uncertainty on the
inputs (easily achieved using Monte Carlo draws), but also the uncertainty
in the model itself thanks to the probabilistic nature of our model.
Thus, even for an input with perfect measurement error (i.e. none), our
forecasting model will still return a probability distribution for the
forecasted quantity, due to (i) our measurement uncertainty in the 
hyper-parameters describing the model; and (ii) the intrinsic variability
seen in nature herself around the imposed model.

To enable the community to make use of this, we have written a 
{\tt Python} package, \href{\packagelink}{\packagename} 
\footnote{https://github.com/chenjj2/forecaster} (MIT license), 
which allows a user
to input a mass (or radius) posterior and return a radius (or mass) forecasted
distribution. Alternatively, one can simply input a mean and standard deviation
of mass (or radius), and the package will return an forecasted mean and 
standard deviation of the radius (or mass), This code works for any object with
mass in the range of [$3\times 10^{-4}$ \mearth, $3\times10^5$ \mearth (0.87
\msun)], or [0.1 \rearth, 100 \rearth (9 \rjup)].

We present the details of how we use the MR relation we obtained to
forecast one quantity from the other below.

\subsection{Forecasting Radius}

Predicting radius given mass is straightforward from our model. If the input is
the mean and standard deviation of mass, \href{\packagelink}{\packagename} will
first generate a vector of masses, $\marr$, following a normal distribution 
truncated within the mass range. Otherwise, the code will accept the input mass 
posterior as $\marr$. \href{\packagelink}{\packagename} will then randomly chose 
$n$ realizations of the hyper parameters from the hyper posteriors derived in this
work. A radius will be drawn for each $M \ith$ with each set of hyper parameters 
$\hyper \ith$, as

\begin{equation}
R \ith \sim \normal ( f(M \ith, \hyper \ith), \scatter  \ith).
\label{rgivenm}
\end{equation}

The output in this case is a vector of radius $\rarr$.
It is worth pointing out that since our model uses a Gaussian distribution, 
it is possible that the predicted radius for a given mass turns out to be so small
that no current physical composition model can explain. 
However we choose not to truncate the prediction with any theoretical model and
let our code users to choose what's suitable for them.

\subsection{Forecasting Mass}

Mass cannot be directly sampled given $\rarr$ with our model. To sample mass, 
\href{\packagelink}{\packagename} first creates a grid of mass as $\mgrid$ in the whole mass range 
of our model. Similarly, then we randomly chose $n$ sets of hyper parameters
from the hyper posteriors of our model. For each radius $R \ith$, 
\href{\packagelink}{\packagename} calculates the probability $\pgrid$ of 
$R \ith$ given $M \jth$ with $\hyper \ith$. Finally, 
\href{\packagelink}{\packagename} samples $M \ith$ from $\mgrid$ with 
$\pgrid$. The output in this case is a vector of mass $\marr$.

\subsection{Examples: Kepler-186f and Kepler-452b}

An illustrative example of \href{\packagelink}{\packagename} in action, we here
forecast the masses of arguably the two most Earth-like planets discovered by
\textit{Kepler}, Kepler-186f and Kepler-452b.

\begin{figure*}
\begin{center}
\includegraphics[width=18.0cm,angle=0,clip=true]{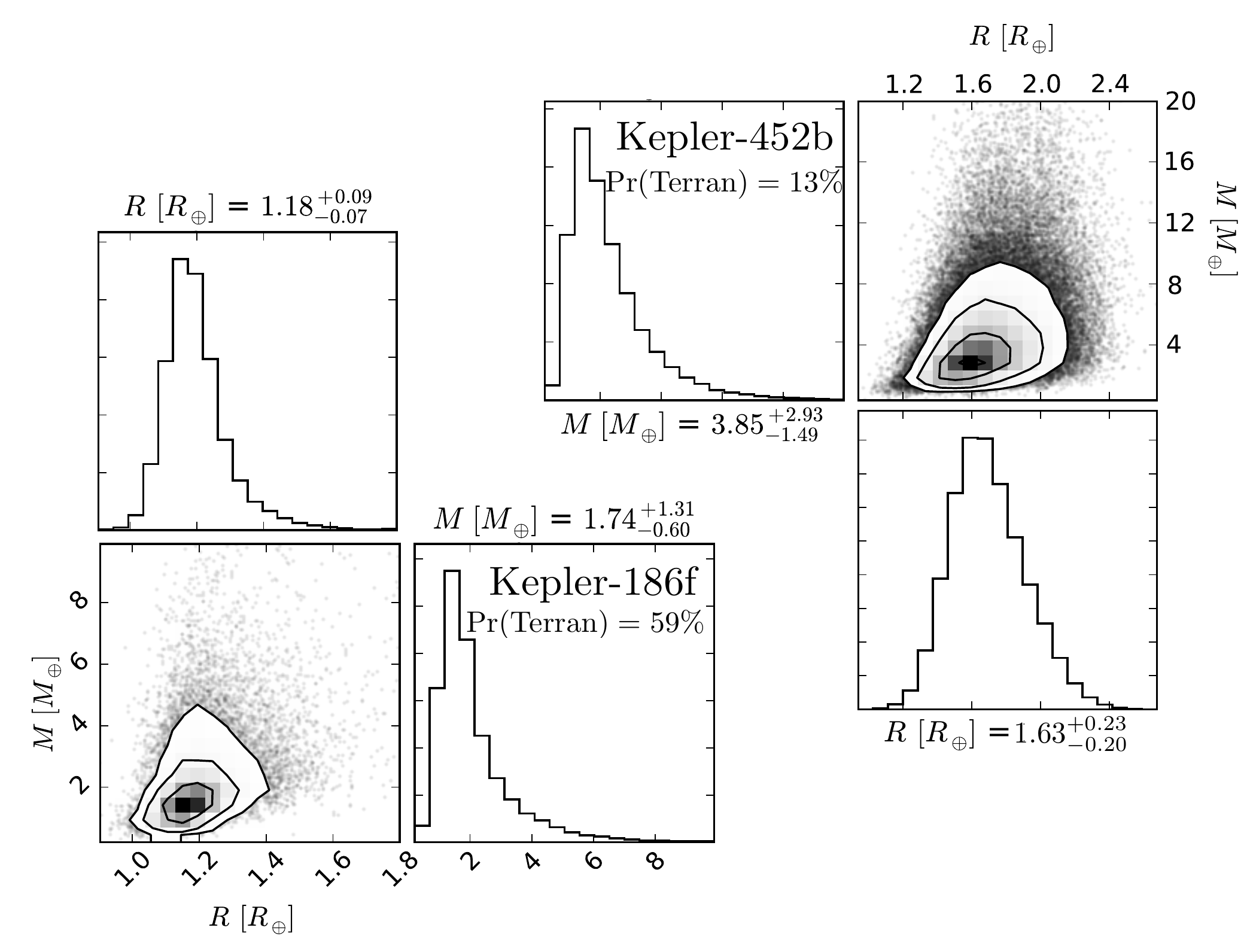}
\caption{
Posterior distributions of the radius (measured) and mass (forecasted)
of two habitable-zone small planets, with predictions produced by
\href{\packagelink}{\packagename} (triangle plots generated using
\href{https://github.com/dfm/corner.py.git}{corner.py}).
}
\label{fig:example}
\end{center}
\end{figure*}

Kepler-186f was discovered by \citet{2014Sci...344..277Q}, reported to
be a $(1.11\pm0.14)$\,\rearth\ and receiving $32_{-4}^{+6}$\% the insolation
received by the Earth. A re-analysis by \citet{2015ApJ...800...99T} refined 
the radius to $(1.17\pm0.08)$\,\rearth\ and we use the radius posterior samples
from that work as our input to \href{\packagelink}{\packagename}. As shown
in Figure~\ref{fig:example}, we predict a mass of
$1.74_{-0.60}^{+1.31}$\,\mearth, with 59\% of the samples lying within the
\terrs\ classification. Therefore, in agreement with the discover paper of 
\citet{2014Sci...344..277Q}, we also predict that Kepler-186f is most
likely a rocky planet.

Kepler-452b was discovered by \citet{2015AJ....150...56J} and was found to have
a very similar insolation to that of the Earth, differing by a factor if just
$1.10_{-0.22}^{+0.29}$. The reported radius of $1.63_{-0.20}^{+0.23}$\,\rearth\
means that Kepler-452b would be unlikely to be rocky using the definition
resulting from the analysis of \citep{2015ApJ...801...41R}. Using the reported radius with 
\href{\packagelink}{\packagename} predicts that $M=3.9_{-1.5}^{+2.9}$\,\mearth,
with only 13\% of samples lying within the \terr\ worlds classification (see 
Figure~\ref{fig:example}). Therefore, in contrast to the discovery paper of 
\citet{2015AJ....150...56J}, we predict that Kepler-452b is unlikely to be a 
rocky planet.

\section{DISCUSSION}
\label{sec:discussion}

In this work, we have developed a new package, called
\href{\packagelink}{\packagename}, to predict the mass (or radius) of an
object based upon the radius (or mass). Our code uses a new probabilistic
mass-radius relation which has been conditioned upon the masses of radii of 
\Ndata\ objects spanning dwarf planets to late-type stars. Aside from
enabling forecasting, this exercise naturally performs classification of the
observed population, since we fit for the transitional points. Since the
observed population has been classified in this way, future objects can
also be probabilistically classified too, which is another feature of 
\href{\packagelink}{\packagename}.

As discussed in Section~\ref{sec:intro}, expected applications may include a newly
discovered transiting planet candidate could have its mass forecasted in order
to estimate the detectability with radial velocities. Vice versa, a newly 
discovered planet found via radial velocities may be considered for transit 
follow-up and our code can predict the detectability given the present 
constraints. Another example might be to forecast the scale height of a
small planet found by \TESS\ for atmospheric followup with \JWST, where
\href{\packagelink}{\packagename} would also calculate the probability of the
object being a \terr\ world.

The classification aspect of our work, which is essentially a free by-product
of our approach, provides some interesting insights:

\begin{itemize}
\item[{\small$\blacktriangleright$}]
There is no discernible change in the MR relation from Jupiter to brown dwarfs.
Brown dwarfs are merely high-mass planets, when classified using their size 
and mass.
\item[{\small$\blacktriangleright$}]
There is no discernible change in the MR relation from dwarf planets to the
Earth. Dwarf planets are merely low-mass planets, when classified using their
size and mass.
\item[{\small$\blacktriangleright$}]
The transition from \nepts\ to \jovis\ occurs at $0.414_{-0.065}^{-0.057}$\,
\mjup, meaning that Saturn is close to being the largest occuring \nept\ world.
This is the first empirical inference of this divide.
\item[{\small$\blacktriangleright$}]
The transition from \terrs\ to \jovis\ occurs at $2.04_{-0.59}^{+0.66}$\,
\mearth, meaning that the Earth is close to being the largest occuring 
solid world. Rocky ``Super-Earths'', then, can be argued to be a fictional 
category.
\end{itemize}

This latter point may seem remarkable given that ``Super-Earths'' have become
part of the astronomical lexicon. The large number of 2-10\,\mearth\ planets
discovered is often cited as evidence that Super-Earths are very common and
thus Solar System's makeup is unusual \citep{2013AREPS..41..469H}. However, if
the boundary between \terr\ and \nept\ worlds is shifted down to 2\,\mearth, 
the Solar System is no longer unusual. Indeed, by our definition three of the 
eight Solar System planets are \nept\ worlds, which are the most common type of
planet around other Sun-like stars \citep{2014ApJ...795...64F}.

As shown earlier, whilst our value is lower than previous estimates, it 
is fully compatible with previous estimates from both theory (e.g. see 
\citealt{2014ApJ...792....1L}) and independent population studies (e.g. see 
\citealt{2015ApJ...801...41R}). The uncertainty on our inference of this key
transition is large ($\sim33$\%) due to the paucity of objects with 
$>3$\,$\sigma$ precision masses and radii in the Earth-mass regime. Future work
could hopefully improve the precision to $\sim10$\% by using a larger sample, 
which will inevitably be found in the coming years, or by extending our
method to include lossier measurements and upper limits by directly re-fitting
the original observations (which was beyond the scope of this work). In any
case, these divides are unlikely sharp, with counter-examples such as the
\mearth-mass \nept\ world KOI-314c \citep{2015ApJ...813...14K}.

A wholly independent line of thinking can also be shown to support the
provocative hypothesis that the divide between \terr\ and \nept\ worlds is
much lower than the canonical 10\,\mearth\ limit. Recently,
\citet{2016MNRAS.456L..59S} made a Bayesian argument using population bias to
infer that inhabited, \terr\ worlds should have radii of $R<1.2$\,\rearth\ to 
95\% confidence. Assuming an Earth-like core-mass fraction, this limit 
corresponds to 2.0\,\mearth\ \citep{2016ApJ...819..127Z}. This is also
compatible with our determination and again argues for effectively a paucity of 
Super-Earths. It may be, then, that the Earth is the Super-Earth we have been 
looking for all along.

%






\clearpage
\appendix


\begin{center}
\begin{longtable}{l|l|l|l}
\caption[Masses and radii used for this study]{
Masses and radii used for this study.} \label{tab:data} \\
\multicolumn{1}{c|}{Name} & \multicolumn{1}{c|}{Mass} & \multicolumn{1}{c|}{Radius} & \multicolumn{1}{c}{Reference} \\ \hline 
\endfirsthead
\multicolumn{4}{c}%
{\tablename\ \thetable{} -- continued from previous page} \\
\multicolumn{1}{c|}{Name} &
\multicolumn{1}{c|}{Mass} &
\multicolumn{1}{c|}{Radius} &
\multicolumn{1}{c}{Reference}  \\ \hline 
\endhead
\hline \multicolumn{4}{|r|}{{Continued on next page}} \\ \hline
\endfoot
\hline \hline
\endlastfoot
NGC6791 KR V20	& $(0.827\pm0.004)$\,\msun & $(0.768\pm0.006)$\,\rsun & \cite{2010A+ARv..18...67T} \\
HD 124784				& $(0.854\pm0.003)$\,\msun & $(0.830\pm0.004)$\,\rsun & \cite{2010A+ARv..18...67T} \\
Parenago 1478		& $(0.727\pm0.010)$\,\msun & $(1.063\pm0.011)$\,\rsun & \cite{2010A+ARv..18...67T} \\
HD 7700					& $(0.764\pm0.004)$\,\msun & $(0.835\pm0.018)$\,\rsun & \cite{2010A+ARv..18...67T} \\
BD+34 4217			& $(0.814\pm0.013)$\,\msun & $(0.838\pm0.011)$\,\rsun & \cite{2010A+ARv..18...67T} \\
TYC 3629-740-1	& $(0.869\pm0.004)$\,\msun & $(0.964\pm0.004)$\,\rsun & \cite{2010A+ARv..18...67T} \\
GU Boo A				& $(0.610\pm0.006)$\,\msun & $(0.627\pm0.016)$\,\rsun & \cite{2010A+ARv..18...67T} \\
GU Boo B				& $(0.600\pm0.006)$\,\msun & $(0.624\pm0.016)$\,\rsun & \cite{2010A+ARv..18...67T} \\
\end{longtable}
\end{center}

\listofchanges

\end{document}